\documentclass[aps,prl,twocolumn,footinbib,groupedaddress,showpacs]{revtex4-1}

\usepackage{graphicx,amsmath}% Include figure files

\begin{document}

%\preprint{}

\title{Vortex Created by Skyrmion Spin Texture under Magnetic Field}

\author{Takao Morinari}
\email[]{morinari@yukawa.kyoto-u.ac.jp}
%\thanks{}
\affiliation{Yukawa Institute for Theoretical Physics, Kyoto
University, Kyoto 606-8502, Japan}

\date{\today}

\begin{abstract}
We show that a vortex current is created around a skyrmion spin texture 
under magnetic field due to a radial spin motive force
in a two-dimensional metal with localized magnetic moments
even in the absence of any superconductivity correlations.
The effect is expected both for ferromagnetic and 
for antiferromagnetic systems.
The Skyrmion-induced vortex mechanism provides
a picture for large Nernst signals observed in 
the pseudogap phase of the high-$T_c$ cuprates.
\end{abstract}

% 72.15.Gd Galvanomagnetic and other magnetotransport effects 
%          (see also 75.47.-m Magnetotransport phenomena; 
%           materials for magnetotransport)
% 75.47.-m Magnetotransport phenomena; materials for magnetotransport
% 72.25.-b Spin polarized transport
% 74.72.-h Cuprate superconductors
%\keywords{}

\maketitle

\newcommand{\be}{\begin{equation}}
\newcommand{\ee}{\end{equation}}
\newcommand{\bea}{\begin{eqnarray}}
\newcommand{\eea}{\end{eqnarray}}
\newcommand{\bsigma}{\boldsymbol{\sigma}}

%\section{Introduction}
%\subsection{Skyrmion}
Skyrmions were originally proposed to describe baryons
in terms of meson fields \cite{Skyrme61,*Skyrme62}.
%made out of meson fields \cite{Skyrme62}.
The central idea of constructing a topological object starting from
an underlying field theory is widely used in many areas 
of physics \cite{MultifacetedSkyrmion}.
In condensed matter physics 
the existence of skyrmions is suggested
and confirmed in part
in quantum Hall systems with filling fraction $\nu \simeq 1$ 
\cite{Sondhi93,Barrett95},
spinor Bose-Einstein condensate \cite{Leslie09},
itinerant ferromagnets \cite{Rossler06,*Binz06,*Tewari06},
nematic liquid crystals \cite{Bogdanov03},
and high-temperature superconductors 
\cite{Shraiman90,*Gooding91,*Haas96,*MorinariMFS,*Pereira07,Raicevic10}.
%Interestingly 
From the topological nature of skyrmions
the Berry phase effect \cite{Berry84} leads to
an effective magnetic field,
which has been studied intensively 
in various systems\cite{[{See, for a review, }]Nagaosa10AHE}.
%Such a field affects Hall effect and experimentally it was reported
%in MnSi that there is a component associated with skyrmions 
%\cite{Lee09,Neubauer09,Timm09}.
In this Letter I show that under magnetic field 
skyrmions create a vortex current 
due to spin motive force
without relying on any superconductivity correlations.
The skyrmion-induced vortex mechanism is applicable
not only to ferromagnetic systems but also to antiferromagnetic systems.

%a well known example
%is the quantum Hall system with the filling fraction $\nu \sim 1$ \cite{Sondhi93}. 
%Projecting electron fields into the lowest Landau level
%leads to a non-linear $\sigma$ model with a Coulomb repulsion
%between field configurations, and the existence of skrymions
%was confirmed by the Knight shift measurement \cite{Barrett95}.
%The presence of skyrmions is also suggested in cold atom systems,

In a two-dimensional isotropic ferromagnet, 
where the local magnetization direction vector 
is represented by $\mathbf{m} = (m_1,m_2,m_3)$,
a skyrmion spin texture at the origin has the following form,
%as schematically shown in Fig.~\ref{fig_skyrmion}(a)
\be
{\mathbf{m}}\left( {\mathbf{r}} \right) = \left( {\frac{{2\lambda x}}
{{r^2  + \lambda ^2 }},\frac{{2\lambda y}}
{{r^2  + \lambda ^2 }},\frac{{r^2  - \lambda ^2 }}
{{r^2  + \lambda ^2 }}} \right),
\label{eq_sky}
\ee
where $r=\sqrt{x^2 + y^2}$ and $\lambda$ is the core size of the skyrmion.
At infinity $\mathbf{m}$ approaches to
the magnetization direction, which is taken as $\hat{e}_3$.
%$\mathbf{m} \rightarrow \hat{e}_3$.
(Hereafter $\hat{e}_j$ represents the unit vector in the $j$-axis.)
This skyrmion spin texture is characterized by the topological charge,
\be
Q = \frac{1}{4\pi}
\int {d^2 {\mathbf{r}}} \,{\mathbf{m}}\left( {\mathbf{r}} \right) 
\cdot \left[ {\partial _x  {\mathbf{m}}\left( {\mathbf{r}} \right) 
\times \partial _y  {\mathbf{m}}\left( {\mathbf{r}} \right)} \right].
\ee
If conduction electrons interact with $\mathbf{m}$
through an exchange coupling
\be
\mathcal{H}_c  = MJ_c \int {d^2 {\mathbf{r}}} 
\left[ {\psi ^\dag  \left( {\mathbf{r}} \right)
\bsigma \psi \left( {\mathbf{r}} \right)} \right] 
\cdot {\mathbf{m}}\left( {\mathbf{r}} \right),
\ee
where $M$ is the magnetization and 
the vector $\bsigma$ is $\bsigma = (\sigma_1,\sigma_2,\sigma_3)$
with $\sigma_j$ the Pauli matrices,
the Berry phase effect associated with
the non-vanishing scalar chirality density 
${\mathbf{m}}\left( {\mathbf{r}} \right) \cdot 
\left[ {\partial _x {\mathbf{m}}\left( {\mathbf{r}} \right) 
\times \partial _y {\mathbf{m}}\left( {\mathbf{r}} \right)} \right]$
produces an effective magnetic field effect for the conduction electrons.
Anomalous Hall effect arising from such a Berry phase effect
has been discussed in manganites \cite{Ye99,*Matl98}, 
pyrochlores \cite{Taguchi01}, 
and MnSi\cite{Lee09,*Neubauer09,*Timm09}.
%Experimentally the effective magnetic field effect 
%is not directly observable because it is not a real magnetic field.
%So usually it is difficult to verify the presence of skyrmions through
%this effective magnetic field.

In this Letter, we study another Berry phase effect
that emerges under magnetic field.
The effect is associated with the spin motive force
discussed in the context of spintronics \cite{Barnes07} and
experimentally observed recently \cite{Hai09,*Yang09}.
%Applying magnetic field leads to spin precession.
If there is a domain wall, the spin motive force is induced
along the domain wall under magnetic field.
Here the same analysis is applied to skyrmion spin textures.

%\section{Model and Formalism}
% simple_skyrmion/note/spin_currents_around_skyrmion.doc
We start with a ferromagnetic metal,
%\bea
\be
%\mathcal{H} &=& \int {d^2 {\mathbf{r}}} 
\mathcal{H} = \int {d^2 {\mathbf{r}}} 
%\left\{ 
\sum\limits_{s=\uparrow,\downarrow}
{\psi _s ^\dag  \left( {\mathbf{r}} \right)
\widehat{K}\left( { - i\hbar \nabla } \right)\psi _s \left( {\mathbf{r}} \right)}
%\right. 
%\nonumber \\ 
%& & 
%\left. 
%+ MJ_c \left[ {\psi ^\dag  \left( {\mathbf{r}} \right) \bsigma \psi 
%\left( {\mathbf{r}} \right)} \right] \cdot {\mathbf{m}}\left( {\mathbf{r}} \right) 
+ \mathcal{H}_c
%\right\}
+ \mathcal{H}_{\bf m},
%\eea
\ee
where ${\hat K}$ is the kinetic energy operator.
%and $J_c$ is the coupling between the conduction electron spin and 
%the local magnetization direction vector $\mathbf{m} = (m_1,m_2,m_3)$.
%The magnetization of the system is given by $M \int d^2 {\bf r} {\bf m}({\bf r})$.
The term $\mathcal{H}_{\bf m}$ describes the interaction for $\mathbf{m}$,
and contains Dzyloshinsky-Moriya interactions.
Here we assume that $\mathcal{H}_{\bf m}$ stabilizes 
a skyrmion texture for $\mathbf{m}$,
and do not discuss explicit mechanisms 
for stabilizing a skrymion 
\footnote{See Ref.~\cite{Rossler06} for discussions
about $\mathcal{H}_{\bf m}$.}.

The coupling term $\mathcal{H}_c$ tends to align
the conduction electron spin to $\mathbf{m}$.
In order to include this correlation effect,
we perform an SU(2) gauge transformation, 
$\psi ({\bf r})  \to U ({\bf r}) \psi ({\bf r})$,
where
$U\left( {\mathbf{r}} \right) = {\mathbf{n}}\left( {\mathbf{r}} \right) \cdot \bsigma$
\cite{Korenman77,*Volovik87,Tserkovnyak08}
with 
\be
{\mathbf{n}}\left( {\mathbf{r}} \right) = 
\frac{{\left( {m_1 \left( {\mathbf{r}} \right),m_2 \left( {\mathbf{r}} \right),1 + m_3 
\left( {\mathbf{r}} \right)} \right)}}
{{\sqrt {2\left[ {1 + m_3 \left( {\mathbf{r}} \right)} \right]} }}.
\ee
The gauge potential associated with the SU(2) gauge transformation is
given by ${\mathbf{a}} =  - i\hbar U^\dag  \nabla U$ and
$a_0  = i\hbar U^\dag  \partial _t U$.
In terms of ${\bf n}$, we obtain
\be
{\mathbf{a}} = \hbar \bsigma  \cdot 
%\left[ 
\left( 
{\mathbf{n}}
%\left( {\mathbf{r}} \right) 
\times \nabla {\mathbf{n}}
%\left( {\mathbf{r}} \right) 
%\right],
\right),
\ee
\be
a_0  =  - \hbar \bsigma  \cdot 
%\left[ 
\left(
{\mathbf{n}}
%\left( {\mathbf{r}} \right) 
\times 
\partial _t {\mathbf{n}}
%\left( {\mathbf{r}} \right)
%\right].
\right).
\ee
Note that $\partial_t {\bf n}$ is computed from 
the Heisenberg equation of motion.
From these expressions we find the SU(2) gauge field,
\be
{\mathbf{e}} =  - 2\hbar \bsigma  \cdot 
%\left[ 
\left(
{\partial _t {\mathbf{n}}
%\left( {\mathbf{r}} \right) 
\times \nabla {\mathbf{n}}
%\left( {\mathbf{r}} \right)
} 
%\right],
\right),
\ee
\be
{\mathbf{b}} = \hbar \varepsilon _{ijk} \sigma_i 
\left( {\nabla n_j } \right) \times \left( {\nabla n_k } \right),
\ee
where $\varepsilon_{ijk}$ is the antisymmetric tensor.

Now we calculate the SU(2) gauge field created by
a skrymion at the origin, Eq.~(\ref{eq_sky}).
%We have assumed that the magnetization is in $\hat{e}_3$ at infinity.
Introducing the cylindrical coordinate $(r,\phi,z)$, we obtain
\be
{\mathbf{b}} = - \frac{{2\hbar \lambda ^2 }}
{{r\left( {r^2  + \lambda ^2 } \right)^2 }}\left( {\begin{array}{*{20}c}
   r & {\lambda e^{- i\phi } }  \\
   {\lambda e^{ i\phi } } & -r  \\
 \end{array} } \right) \hat{e}_z,
\label{eq_b}
\ee
and ${\bf e}=0$.
It is easy to see that
$b_{zu} \equiv [b_z]_{\uparrow \uparrow}$ is equal to 
${\mathbf{m}} \cdot 
\left( {\partial _x {\mathbf{m}} \times \partial _y {\mathbf{m}}} \right)/2$
\cite{Ye99}.
The field $b_{zu}$ behaves like a magnetic field for 
the conduction electrons.
For $\lambda=100 \AA$,
the average field strength is 
$\int_0^{\lambda} dr 2\pi r 
%[b_z]_{\uparrow \uparrow}
b_{zu}/(2\pi e \lambda^2) \simeq 3.3$~T.
With decreasing $\lambda$, $b_{zu}$ increases,
and so $b_{zu}$ can be very large.
However, this effective field is not directly observable
because it is not a real magnetic field.

Now we apply the magnetic field ${\bf B} = (0,0,-B)$ to the system.
From the Zeeman energy term the dynamics of ${\bf m}$ is given by
\be
\frac{\partial }
{{\partial t}}{\mathbf{m}} = \alpha {\mathbf{B}} \times {\mathbf{m}},
\ee
where $\alpha = g M\mu_B/\hbar$
with $g$ the g-factor and $\mu_B$ the Bohr magneton.
Using the solution of this equation of motion, we find
\be
{\mathbf{n}}\left( {{\mathbf{r}},t} \right) 
= \frac{{\lambda \cos \Phi \left( t \right)  \hat{e}_1  
+ \lambda \sin \Phi \left( t \right)\hat{e}_2  + r \hat{e}_3 }}
{{\sqrt {r^2  + \lambda ^2 } }},
\ee
with $\Phi \left( t \right) = \phi -\alpha Bt + \phi _0$.
($\phi_0$ is a constant.)
The field ${\bf b}$ is given by Eq.~(\ref{eq_b}) with
$\phi$ being replaced by $\Phi(t)$.
Now the field ${\bf e}$ is nonzero and given by
\be
{\mathbf{e}} =  \frac{{2\alpha B\hbar \lambda ^2 }}
{{\left( {r^2  + \lambda ^2 } \right)^2 }}\left( {\begin{array}{*{20}c}
   r & {\lambda e^{ - i\Phi(t) } }  \\
   {\lambda e^{i\Phi(t) } } & { - r}  \\

 \end{array} } \right)\hat{e}_r 
\ee
%The fields $b_{zu}$
%$[b_z]_{\uparrow \uparrow}$ 
%and $e_{zu} \equiv [e_z]_{\uparrow \uparrow}$
%as a function of $r$ is shown in Fig.~\ref{fig_fields}.
Note that the effective field ${\bf e}$ is created 
in the radial direction 
\cite{[{A similar calculation was presented in }]Wong09}.
%\cite{Wong09}.
This is understood as follows.
If one sees the skyrmion spin texture along a straight line 
crossing the origin in the x-y plane,
the spin configuration looks like a domain wall.
From the analysis of the domain wall we know that
the spin motive force is created along it \cite{Barnes07}.
The same is true in each direction.
Thus, the spin motive force ${\bf e}$ is created in the radial direction
for skyrmion spin textures.
%=====================================================================
% codes/skyrmion1_0.pov 
%     i)   bmp
%     (  Illustrator: fig_skyrmion_ai.jpg -> EPS-CONV )
%     ii)   Inkscape
%     iii)  EPS
%\begin{figure}
%   \begin{center}
%    %\includegraphics[width=0.8 \linewidth]{fig_skyrmion2cmp.eps}
%    %\includegraphics[width=0.8 \linewidth]{fig_skyrmion2.eps}
%    % fig_skyrmion3.eps (Inkscape)
%    %\includegraphics[width=0.8 \linewidth]{fig_skyrmion3.eps}
%    %%%%%%%%%%%%%%%%%%%%%%%%%%%%%%%%%%%%%%%%%%%%%%%%%%%%%%%%%%%%%%%%%%%
%    % note/magnetic_field_induced.doc
%    % codes/gpf_bzez2.plt
%    %%%%%%%%%%%%%%%%%%%%%%%%%%%%%%%%%%%%%%%%%%%%%%%%
%    % (codes/gpf_bzezBz2.plt)
%    % codes/bzf1_0.cc
%    % (codes/gpf_bzezBz.plt)
%    %\includegraphics[width=0.8 \linewidth]{fig_bzezBz.eps}
%    %\includegraphics[width=0.8 \linewidth]{fig_bzezBz2.eps}
%    \includegraphics[width=0.8 \linewidth]{fig_bzez2.eps}
%    %%%%%%%%%%%%%%%%%%%%%%%%%%%%%%%%%%%%%%%%%%%%%%%%%%%%%%%%%%%%%%%%%%%
%   \end{center}
%   \caption{ \label{fig_fields}
%	(color online) 
%	The fields 
%        %$[e_z]_{\uparrow \uparrow} / (2\alpha B \hbar /\lambda)$ and 
%	%$[b_z]_{\uparrow \uparrow}/ (2 \hbar/\lambda^2)$
%        $e_{zu}/ (2\alpha B \hbar /\lambda)$ and 
%	$b_{zu}/ (2 \hbar/\lambda^2)$
%	as a function of $r/\lambda$.
%	%The skyrmion spin texture \ref{eq_sky} with $\lambda=4 a$
%	%on the square lattice with the lattice constant $a$.
%    }
% \end{figure}
%=====================================================================

\begin{figure}
   \begin{center}
    \includegraphics[width=0.8 \linewidth]{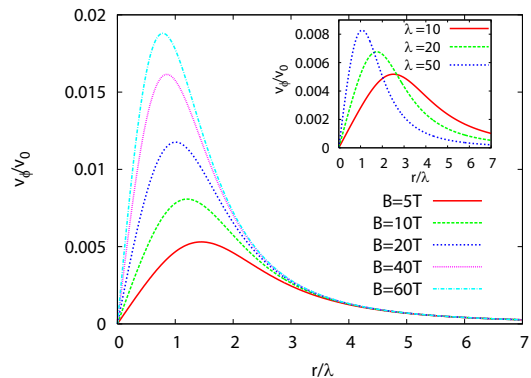}
   \end{center}
   \caption{ \label{fig_drift}
     (color online)
	The drift velocity as a function of $r/\lambda$ for 
	different magnetic fields at $\lambda=40 \AA$ with
	$v_0 = gMc$ with $c$ the speed of light.
	The inset shows the drift velocity versus $r/\lambda$ 
	for different $\lambda$ (in units of $\AA$ ) at $B=10$~T.
    }
 \end{figure}

%\subsection{Drift velocity}
Since $e_{zu}$ is non-zero and there is the external magnetic field
and $b_{zu}$, a drift of conduction electrons is induced.
The drift is circular around the skyrmion because $e_{zu}$ is
in the radial direction.
Thus, the drift motion leads to the vortex current.
The drift velocity is
\be
v_\phi   =  \frac{{2\alpha B\lambda ^2 r}}
{{2\lambda ^2  + 
%\frac{{eB}}{\hbar }
(eB/\hbar)
\left( {r^2  + \lambda ^2 } \right)^2 }}.
\label{eq_v}
\ee
Figure \ref{fig_drift} shows this drift velocity 
as a function of $r/\lambda$.
For $r/\lambda < 1$, $b_{zu}$ plays a major role
for the drift.
While for $r/\lambda > 1$, $B$ plays a major role.
For $B>0$ the external magnetic field is parallel 
to $b_{zu}$.
%$[b_z]_{\uparrow \uparrow}$.
If we consider an anti-skyrmion,
the drift velocity changes its direction
at $r = \sqrt {\lambda \left( {\sqrt 2 \ell_B  - \lambda } \right)}$
with $\ell_B = \sqrt{eB/\hbar}$ the magnetic length.
We do not consider this case in the following analysis
because such a current distribution
is energetically unfavorable, though it is not forbidden.

%=====================================================================
% codes/dv1_0.cc -> gpf_dv1_0.plt
%      Illustrator -> ghostview (File -> "PS to EPS")
% note/spin_current_around_skyrmion.doc
%	Spin precession under magnetic field/Drift velocity functional form
%
%=====================================================================

\begin{figure}
   \begin{center}
    \includegraphics[width=0.8 \linewidth]{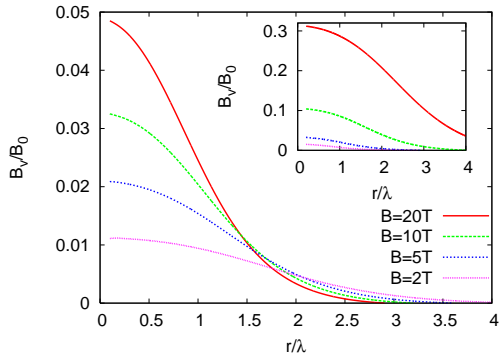}
   \end{center}
   \caption{ \label{fig_vortex_current}
	(color online)
	The magnetic field $B_v$ created by the vortex current
	for different magnetic field at $z=0$ with $\lambda=40 \AA$.
	The inset shows $B_v$ for $\lambda = 10,20,40,60$
	(in units of $\AA$) from top to bottom.
    }
 \end{figure}

%\frac{{\left[ {\mathbf{e}} \right]_{ \uparrow  \uparrow }  \times B\hat{e}_z }}
%{e{B^2 }} = g M \frac{\hbar }
%{m}\frac{{\lambda ^2 r}}
%{{\left( {r^2  + \lambda ^2 } \right)^2 }}\hat{e}_\phi.
%\ee
The circular drift velocity (\ref{eq_v}) leads to a vortex current.
The magnetic field $B_v$ created by this vortex current 
is computed as in that for vortex current in a layered superconductor \cite{Clem91}:
\be
B_{v}  = \int_0^\infty  {dq} q
J_0 \left( {q\rho } \right)A_0 \left( q \right)
\exp \left( { - q\left| z \right|} \right),
\label{eq_Bv}
\ee
\be
A_0 \left( q \right) =  - \frac{{2\pi }}
{c}\int_0^\infty  {d\rho } \rho J_1 \left( {q\rho } \right)
\left( {{\mathbf{j}}_{\rm 2d} } \right)_\phi,
\label{eq_A0}
\ee
with $J_n(x)$ the Bessel function and 
the two-dimensional current density is
$\left( {{\mathbf{j}}_{\rm 2d} } \right)_\phi   = e n_{\rm 2d} v_\phi$.
Here $n_{\rm 2d}$ is the two-dimensional conduction electron density.
This magnetic field $B_v$ is shown 
in Fig.~\ref{fig_vortex_current}.
The field is measured in units of tesla and 
is normalized by a dimensionless parameter
$B_0 = g M n_{\rm 2d} \lambda^2$.
%It is worth pointing out that generally 
Note that usually vortex currents are induced
if there is a superconductivity correlation.
However, the skyrmion-induced vortex current does not require 
any superconductivity correlations.
%Furthermore, the magnetic field created by vortex currents 
%is the real magnetic field.
%Therefore, it is experimentally observable.

%=====================================================================
% codes/vc2_0.cc, vc2_2.cc
%       gpf_vc2_0b.plt (gpf_vc2_0.plt), PS to EPS
%       fig_vc2_0ill.ai
% CygwinData/Data0702_10
%=====================================================================

\begin{figure}
   \begin{center}
    \includegraphics[width=0.8 \linewidth]{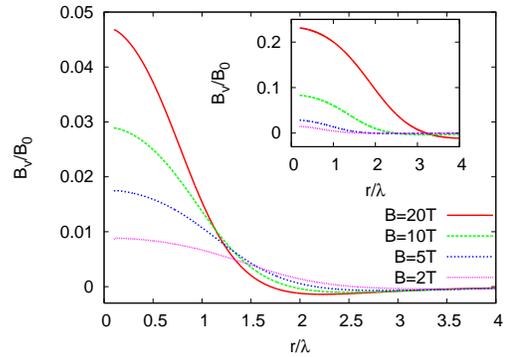}
   \end{center}
   \caption{ \label{fig_af}
	(color online)
	The magnetic field $B_v$ as a function of $r/\lambda$ 
        for the antiferromagnetic case.
        The value of $B_v$ is computed
	for different magnetic field at $z=0$ with $\lambda=40 \AA$.
	The inset shows $B_v$ for $\lambda = 10,20,40,60$
	(in units of $\AA$) from top to bottom.
    }
 \end{figure}

The same formula can be applied to the antiferromagnetic case as well.
Let us consider a square lattice and introduce A and B sublattice.
For this case we take $\mathbf{m}({\bf r})$ as the local staggered 
magnetization direction vector.
The conduction electron spin residing at ${\bf R}_j$ is aligned to 
$\mathbf{m} ({\bf R}_j)$ by the unitary transformation matrix $U({\bf R}_j)$.
Carrying out this transformation, we obtain the same Hamiltonian
except for the exchange interaction term.
The sign of the exchange interaction term is different for A and B
sublattice.
In the strong coupling limit, the conduction electron 
spin at B sublattice is anti-parallel to that at A sublattice.
The relevant components of the SU(2) fields are
\be
\left[ {\mathbf{e}} \right]_{\uparrow \downarrow }  =  
(\left[ {\mathbf{e}} \right]_{\downarrow \uparrow })^*  =  
\frac{{2\alpha B\hbar \lambda ^3 }}
{{\left( {r^2  + \lambda ^2 } \right)^2 }}e^{ - i\Phi } \hat{e}_r,
\ee
\be
\left[ {\mathbf{b}} \right]_{\uparrow \downarrow }  = 
(\left[ {\mathbf{b}} \right]_{\downarrow \uparrow} )^* = 
- \frac{{2\hbar \lambda ^3 }}
{{r\left( {r^2  + \lambda ^2 } \right)^2 }}e^{- i\Phi } \hat{e}_z.
\ee
Note that the SU(2) gauge fields acting
on the conduction electrons are not staggered,
in spite of the fact that the magnetization direction vector is staggered.
Denoting ${\bf e}=\sum_j {\bf e}_j \sigma_j$
and ${\bf b}=\sum_j {\bf b}_j \sigma_j$,
we find the gauge field components ${\bf e}_j$ and ${\bf b}_j$.
For a bond vector connecting from B sublattice to A sublattice,
these components take opposite sign compared to the case of
a bond vector connecting from A sublattice to B sublattice.
A crucial point is that this sign change cancel completely 
in the drift created by ${\bf e}$ and ${\bf b}$.
The situation is clearer if we carry out an additional 
Unitary transformation 
$V = \exp \left( {i\Phi \sigma _z /2} \right)$
that makes ${\bf e}_{\uparrow \downarrow}={\bf e}_{\downarrow \uparrow}$
and ${\bf b}_{\uparrow \downarrow}={\bf b}_{\downarrow \uparrow}$
and removes the sign change.
The drift velocity calculation is similar to the ferromagnetic case,
and we obtain
\be
v_\phi = 
\frac{{2\alpha B\lambda ^3 r}}
     {{2\lambda ^3  + \left( {eB/\hbar } \right)r\left( {r^2  + \lambda ^2 } \right)^2 }}.
\label{eq_v_af}
\ee
The difference arises from the relevant component of the gauge fields as stated above.
Figure \ref{fig_af} shows 
the magnetic field created by the drift calculated by
Eqs.~(\ref{eq_Bv}) and (\ref{eq_A0}) using Eq.~(\ref{eq_v_af}).
Although $r$ dependence of the effective fields are different between
the antiferromagnetic case and the ferromagnetic case,
the results are similar.
For $r < \lambda$, the effective field ${\bf b}$ plays a major
role for the drift.
In this regime, the drift velocity has the same form as that
for the ferromagnetic case.
This is the reason why we obtained similar results
for both cases for $r<\lambda$.
Meanwhile for $r>\lambda$ the effective electric field 
is suppressed by the factor $\lambda/r$.
Thus, $B_v$ decreases rapidly compared to the ferromagnetic case.

%=====================================================================
%=====================================================================

In order to stabilize the vortex current, we need to keep 
the radial effective electric field ${\bf e}$.
However, in general a radial electric field is screened
by conduction electrons.
Therefore, to stabilize the vortex current 
the Thomas-Fermi length, which is proportional to $1/n_{\rm 2d}$,
should be larger than $\lambda$,
otherwise the effective electric field is screened by the 
conduction electrons.

Now it is natural to ask a question:
What is the difference between the vortex current arising from
the skyrmion spin texture and a conventional vortex
arising from a superconductivity correlation?
As long as the external magnetic field is kept,
the skyrmion-induced vortex current continues to flow
even in the presence of dissipations.
However, the current decays rapidly when one turns off 
the magnetic field.
This makes the crucial difference compared to
conventional vortices in a superconductor.
In addition, the flux quantization occurs only for 
a mesoscopic sample.
% for the skyrmion induced vortex current.

In the formulation above we have assumed that 
there is magnetic long-range order.
But this assumption is not necessary.
Suppose the system does not have magnetic long-range order
and let $\xi$ be the magnetic correlation length.
For $\xi \gg \lambda$, we may apply the formula above 
to the system as well.
If $\xi$ is comparable to $\lambda$, the field ${\bf b}$ and 
${\bf e}$ are multiplied by the factor $\exp (-2 r/\xi)$,
and $M$ should be replaced by the average of spins over the domain.
The amplitude of the fields would decrease by these factors but 
one may still expect a vortex formation.

So far we have assumed that skyrmions are static.
However, it is possible to consider 
a moving skyrmion
\cite{[{A numerical simulation 
of a moving spin texture 
based on the Landau-Lifshitz-Gilbert equation 
was carried out in 
}]
[{ under inplane magnetic field}]
Ohe09} 
and compute a vortex current around it.
For a moving skyrmion, the antiferromagnetic case is much better than
the ferromagnetic case.
If one adopts the non-linear $\sigma$ model
for the description of the skyrmion spin texture,
the skyrmion is a soliton solution of the non-linear $\sigma$ model.
However, for the ferromagnetic case the skyrmion spin texture 
is stable only for the static case.
Because of the quadratic dispersion of spin waves, 
the chiral nature of the skyrmion spin texture is lost 
upon skyrmion propagation.
By contrast, for the antiferromagnetic case
a moving skyrmion spin texture solution is constructed
by a Lorentz boost.
This is possible because the non-linear $\sigma$ model has 
a relativistic form.  
Of course there are deviations from the relativistic dynamics
in the real system,
and those deviations would lead to a finite life time
even in the antiferromagnetic case.
%(The computation of the life time requires information about the model.)

In order to observe the vortex currents predicted above
candidate systems are Fe$_{1-x}$Co$_x$Si, 
where real-space observation of a two-dimensional skyrmion crystal 
was reported \cite{Yu10}, and MnSi, where magnetoresistance measurements 
suggest the presence of skyrmions \cite{Lee09,Neubauer09}.
Another interesting application of the theory is that for
cuprate high-temperature superconductors.
In recent experiments large Nernst effects were observed\cite{Wang06}
above the superconductivity transition temperature $T_c$.
For the single band system, which is believed to be the case
for the high-$T_c$, the Nernst effect is absent because 
of Sondheimer cancellation\cite{Wang06}.
One scenario for the large Nernst effect above $T_c$
is to assume the presence of preformed Cooper pairs.
Assuming a skyrmion texture carried by a doped hole
provides an alternative scenario.
To make clear whether the skyrmion spin texture is 
formed in the high-$T_c$ cuprates or not,
the best way would be to investigate Li-doped underdoped samples.
Establishing a skyrmion spin texture by observing 
the vortex current around a doped hole bound to a Li ion provides
a crucial step to uncover the mechanism of cuprate high-temperature
superconductivity
\footnote{
Recent magnetization and magnetoresistance measurements\cite{Raicevic10}
suggested the presence of skyrmions
in insulating antiferromagnet La$_2$Cu$_{0.97}$Li$_{0.03}$O$_4$.}.
%%%%%%%%%%%%%%%%%%
%\cite{[{Recent magnetization and magnetoresistance measurements
%in insulating antiferromagnet La$_2$Cu$_{0.97}$Li$_{0.03}$O$_4$ 
%reported in }]
%[{ suggested the presence of skyrmions}]
%Raicevic10}

%\section{Summary}
To conclude, we have discussed a mechanism of vortex current formation 
under magnetic field based on the skyrmion spin texture without 
relying on any superconductivity correlations.
The vortex current is created for antiferromagnetic systems as well as 
ferromagnetic systems.
The presence of the vortex can be verified by observing the magnetic
field created by the vortex current.
Searching for a skyrmion-induced vortex current in underdoped
cuprates would be a key step to justify the relevance of the skyrmion
spin texture in high-$T_c$.

% If in two-column mode, this environment will change to single-column
%\begin{widetext}
%\end{widetext}

% Surround figure environment with turnpage environment for landscape
% figure
% \begin{figure}
% \end{figure}

\begin{acknowledgments}
%This work was supported by KAKENHI (21740252),
%the Grant-in-Aid for Scientific Research from the Ministry of Education, 
%Culture, Sports, Science and Technology (MEXT) of Japan, 
I would like to thank M.~Sigrist for helpful discussions.
This work was supported by 
the Global COE Program "The Next Generation of Physics, 
Spun from Universality and Emergence," 
and Yukawa International Program for Quark-Hadron Sciences at YITP.
\end{acknowledgments}

\bibliography{../../../references/tm_library2}

%Merlin.mbs v4.21 2009-07-09.
\begin{thebibliography}{10}%
\makeatletter
\providecommand \@ifxundefined [1]{%
 \ifx #1\undefined \expandafter \@firstoftwo
 \else \expandafter \@secondoftwo
\fi
}%
\providecommand \@ifnum [1]{%
 \ifnum #1\expandafter \@firstoftwo
 \else \expandafter \@secondoftwo
\fi
}%
\providecommand \enquote [1]{``#1''}%
\providecommand \bibnamefont  [1]{#1}%
\providecommand \bibfnamefont [1]{#1}%
\providecommand \citenamefont [1]{#1}%
\providecommand\href[0]{\@sanitize\@href}%
\providecommand\@href[1]{\endgroup\@@startlink{#1}\endgroup\@@href}%
\providecommand\@@href[1]{#1\@@endlink}%
\providecommand \@sanitize [0]{\begingroup\catcode`\&12\catcode`\#12\relax}%
\@ifxundefined \pdfoutput {\@firstoftwo}{%
 \@ifnum{\z@=\pdfoutput}{\@firstoftwo}{\@secondoftwo}%
}{%
 \providecommand\@@startlink[1]{\leavevmode\special{html:<a href="#1">}}%
 \providecommand\@@endlink[0]{\special{html:</a>}}%
}{%
 \providecommand\@@startlink[1]{%
  \leavevmode
  \pdfstartlink
   attr{/Border[0 0 1 ]/H/I/C[0 1 1]}%
   user{/Subtype/Link/A<</Type/Action/S/URI/URI(#1)>>}%
  \relax
 }%
 \providecommand\@@endlink[0]{\pdfendlink}%
}%
\providecommand \url  [0]{\begingroup\@sanitize \@url }%
\providecommand \@url [1]{\endgroup\@href {#1}{\urlprefix}}%
\providecommand \urlprefix [0]{URL }%
\providecommand \Eprint[0]{\href }%
\@ifxundefined \urlstyle {%
  \providecommand \doi [1]{doi:\discretionary{}{}{}#1}%
}{%
  \providecommand \doi [0]{doi:\discretionary{}{}{}\begingroup
  \urlstyle{rm}\Url }%
}%
\providecommand \doibase [0]{http://dx.doi.org/}%
\providecommand \Doi[1]{\href{\doibase#1}}%
\providecommand \bibAnnote [3]{%
  \BibitemShut{#1}%
  \begin{quotation}\noindent
    \textsc{Key:}\ #2\\\textsc{Annotation:}\ #3%
  \end{quotation}%
}%
\providecommand \bibAnnoteFile [2]{%
  \IfFileExists{#2}{\bibAnnote {#1} {#2} {\input{#2}}}{}%
}%
\providecommand \typeout [0]{\immediate \write \m@ne }%
\providecommand \selectlanguage [0]{\@gobble}%
\providecommand \bibinfo [0]{\@secondoftwo}%
\providecommand \bibfield [0]{\@secondoftwo}%
\providecommand \translation [1]{[#1]}%
\providecommand \BibitemOpen[0]{}%
\providecommand \bibitemStop [0]{}%
\providecommand \bibitemNoStop [0]{.\EOS\space}%
\providecommand \EOS [0]{\spacefactor3000\relax}%
\providecommand \BibitemShut [1]{\csname bibitem#1\endcsname}%
%</preamble>
\bibitem{Skyrme61}%
  \BibitemOpen
  \bibfield{author}{%
  \bibinfo {author} {\bibfnamefont{T.}~\bibnamefont{Skyrme}},\ }%
  \bibfield{journal}{%
  \bibinfo {journal} {Proceedings of the Royal Society of London. Series A,
  Mathematical and Physical Sciences}\ }%
  \textbf{\bibinfo {volume} {260}},\ \bibinfo {pages} {127} (\bibinfo {year}
  {1961})%
  \bibAnnoteFile{NoStop}{Skyrme61}%
\bibitem{Skyrme62}%
  \BibitemOpen
  \bibfield{author}{%
  \bibinfo {author} {\bibfnamefont{T.}~\bibnamefont{Skyrme}},\ }%
  \bibfield{journal}{%
  \bibinfo {journal} {Nucl. Phys}\ }%
  \textbf{\bibinfo {volume} {31}},\ \bibinfo {pages} {556} (\bibinfo {year}
  {1962})%
  \bibAnnoteFile{NoStop}{Skyrme62}%
\bibitem{MultifacetedSkyrmion}%
  \BibitemOpen
  \emph{\bibinfo {title} {The Multifaceted Skyrmion}},\ edited by\ \bibinfo
  {editor} {\bibfnamefont{G.~E.}\ \bibnamefont{Brown}}\ and\ \bibinfo {editor}
  {\bibfnamefont{M.}~\bibnamefont{Rho}}\ (\bibinfo {publisher} {World
  Scientific},\ \bibinfo {year} {2010})%
  \bibAnnoteFile{NoStop}{MultifacetedSkyrmion}%
\bibitem{Sondhi93}%
  \BibitemOpen
  \bibfield{author}{%
  \bibinfo {author} {\bibfnamefont{S.~L.}\ \bibnamefont{Sondhi}}, \bibinfo
  {author} {\bibfnamefont{A.}~\bibnamefont{Karlhede}}, \bibinfo {author}
  {\bibfnamefont{S.~A.}\ \bibnamefont{Kivelson}},\ and\ \bibinfo {author}
  {\bibfnamefont{E.~H.}\ \bibnamefont{Rezayi}},\ }%
  \bibfield{journal}{%
  \bibinfo {journal} {Phys. Rev. B}\ }%
  \textbf{\bibinfo {volume} {47}},\ \bibinfo {pages} {16419} (\bibinfo {year}
  {1993})%
  \bibAnnoteFile{NoStop}{Sondhi93}%
\bibitem{Barrett95}%
  \BibitemOpen
  \bibfield{author}{%
  \bibinfo {author} {\bibfnamefont{S.~E.}\ \bibnamefont{Barrett}}, \bibinfo
  {author} {\bibfnamefont{G.}~\bibnamefont{Dabbagh}}, \bibinfo {author}
  {\bibfnamefont{L.~N.}\ \bibnamefont{Pfeiffer}}, \bibinfo {author}
  {\bibfnamefont{K.~W.}\ \bibnamefont{West}},\ and\ \bibinfo {author}
  {\bibfnamefont{R.}~\bibnamefont{Tycko}},\ }%
  \bibfield{journal}{%
  \bibinfo {journal} {Phys. Rev. Lett.}\ }%
  \textbf{\bibinfo {volume} {74}},\ \bibinfo {pages} {5112} (\bibinfo {year}
  {1995})%
  \bibAnnoteFile{NoStop}{Barrett95}%
\bibitem{Leslie09}%
  \BibitemOpen
  \bibfield{author}{%
  \bibinfo {author} {\bibfnamefont{L.~S.}\ \bibnamefont{Leslie}}, \bibinfo
  {author} {\bibfnamefont{A.}~\bibnamefont{Hansen}}, \bibinfo {author}
  {\bibfnamefont{K.~C.}\ \bibnamefont{Wright}}, \bibinfo {author}
  {\bibfnamefont{B.~M.}\ \bibnamefont{Deutsch}},\ and\ \bibinfo {author}
  {\bibfnamefont{N.~P.}\ \bibnamefont{Bigelow}},\ }%
  \bibfield{journal}{%
  \bibinfo {journal} {Phys. Rev. Lett.}\ }%
  \textbf{\bibinfo {volume} {103}},\ \bibinfo {pages} {250401} (\bibinfo {year}
  {2009})%
  \bibAnnoteFile{NoStop}{Leslie09}%
\bibitem{Rossler06}%
  \BibitemOpen
  \bibfield{author}{%
  \bibinfo {author} {\bibfnamefont{U.}~\bibnamefont{R\"o\ss{}ler}}, \bibinfo
  {author} {\bibfnamefont{A.}~\bibnamefont{Bogdanov}},\ and\ \bibinfo {author}
  {\bibfnamefont{C.}~\bibnamefont{Pfleiderer}},\ }%
  \bibfield{journal}{%
  \bibinfo {journal} {Nature}\ }%
  \textbf{\bibinfo {volume} {442}},\ \bibinfo {pages} {797} (\bibinfo {year}
  {2006})%
  \bibAnnoteFile{NoStop}{Rossler06}%
\bibitem{Binz06}%
  \BibitemOpen
  \bibfield{author}{%
  \bibinfo {author} {\bibfnamefont{B.}~\bibnamefont{Binz}}, \bibinfo {author}
  {\bibfnamefont{A.}~\bibnamefont{Vishwanath}},\ and\ \bibinfo {author}
  {\bibfnamefont{V.}~\bibnamefont{Aji}},\ }%
  \bibfield{journal}{%
  \bibinfo {journal} {Phys. Rev. Lett.}\ }%
  \textbf{\bibinfo {volume} {96}},\ \bibinfo {pages} {207202} (\bibinfo {year}
  {2006})%
  \bibAnnoteFile{NoStop}{Binz06}%
\bibitem{Tewari06}%
  \BibitemOpen
  \bibfield{author}{%
  \bibinfo {author} {\bibfnamefont{S.}~\bibnamefont{Tewari}}, \bibinfo {author}
  {\bibfnamefont{D.}~\bibnamefont{Belitz}},\ and\ \bibinfo {author}
  {\bibfnamefont{T.~R.}\ \bibnamefont{Kirkpatrick}},\ }%
  \bibfield{journal}{%
  \bibinfo {journal} {Phys. Rev. Lett.}\ }%
  \textbf{\bibinfo {volume} {96}},\ \bibinfo {pages} {047207} (\bibinfo {year}
  {2006})%
  \bibAnnoteFile{NoStop}{Tewari06}%
\bibitem{Bogdanov03}%
  \BibitemOpen
  \bibfield{author}{%
  \bibinfo {author} {\bibfnamefont{A.~N.}\ \bibnamefont{Bogdanov}}, \bibinfo
  {author} {\bibfnamefont{U.~K.}\ \bibnamefont{R\"o\ss{}ler}},\ and\ \bibinfo
  {author} {\bibfnamefont{A.~A.}\ \bibnamefont{Shestakov}},\ }%
  \bibfield{journal}{%
  \bibinfo {journal} {Phys. Rev. E}\ }%
  \textbf{\bibinfo {volume} {67}},\ \bibinfo {pages} {016602} (\bibinfo {year}
  {2003})%
  \bibAnnoteFile{NoStop}{Bogdanov03}%
\bibitem{Shraiman90}%
  \BibitemOpen
  \bibfield{author}{%
  \bibinfo {author} {\bibfnamefont{B.~I.}\ \bibnamefont{Shraiman}}\ and\
  \bibinfo {author} {\bibfnamefont{E.~D.}\ \bibnamefont{Siggia}},\ }%
  \bibfield{journal}{%
  \bibinfo {journal} {Phys. Rev. B}\ }%
  \textbf{\bibinfo {volume} {42}},\ \bibinfo {pages} {2485} (\bibinfo {year}
  {1990})%
  \bibAnnoteFile{NoStop}{Shraiman90}%
\bibitem{Gooding91}%
  \BibitemOpen
  \bibfield{author}{%
  \bibinfo {author} {\bibfnamefont{R.~J.}\ \bibnamefont{Gooding}},\ }%
  \bibfield{journal}{%
  \bibinfo {journal} {Phys. Rev. Lett.}\ }%
  \textbf{\bibinfo {volume} {66}},\ \bibinfo {pages} {2266} (\bibinfo {year}
  {1991})%
  \bibAnnoteFile{NoStop}{Gooding91}%
\bibitem{Haas96}%
  \BibitemOpen
  \bibfield{author}{%
  \bibinfo {author} {\bibfnamefont{S.}~\bibnamefont{Haas}}, \bibinfo {author}
  {\bibfnamefont{F.-C.}\ \bibnamefont{Zhang}}, \bibinfo {author}
  {\bibfnamefont{F.}~\bibnamefont{Mila}},\ and\ \bibinfo {author}
  {\bibfnamefont{T.~M.}\ \bibnamefont{Rice}},\ }%
  \bibfield{journal}{%
  \bibinfo {journal} {Phys. Rev. Lett.}\ }%
  \textbf{\bibinfo {volume} {77}},\ \bibinfo {pages} {3021} (\bibinfo {year}
  {1996})%
  \bibAnnoteFile{NoStop}{Haas96}%
\bibitem{MorinariMFS}%
  \BibitemOpen
  \bibfield{author}{%
  \bibinfo {author} {\bibfnamefont{T.}~\bibnamefont{Morinari}},\ }%
  in\ \emph{\bibinfo {booktitle} {The Multifaceted Skrymion}}\ (\bibinfo
  {publisher} {World Scientific},\ \bibinfo {year} {2010})%
  \bibAnnoteFile{NoStop}{MorinariMFS}%
\bibitem{Pereira07}%
  \BibitemOpen
  \bibfield{author}{%
  \bibinfo {author} {\bibfnamefont{A.~R.}\ \bibnamefont{Pereira}}, \bibinfo
  {author} {\bibfnamefont{E.}~\bibnamefont{Ercolessi}},\ and\ \bibinfo {author}
  {\bibfnamefont{A.~S.~T.}\ \bibnamefont{Pires}},\ }%
  \bibfield{journal}{%
  \bibinfo {journal} {J. Phys. Condens. Matter}\ }%
  \textbf{\bibinfo {volume} {19}},\ \bibinfo {pages} {156203} (\bibinfo {year}
  {2007})%
  \bibAnnoteFile{NoStop}{Pereira07}%
\bibitem{Raicevic10}%
  \BibitemOpen
  \bibfield{author}{%
  \bibinfo {author} {\bibfnamefont{I.}~\bibnamefont{Raicevic}}, \bibinfo
  {author} {\bibfnamefont{D.}~\bibnamefont{Popovic}}, \bibinfo {author}
  {\bibfnamefont{C.}~\bibnamefont{Panagopoulos}}, \bibinfo {author}
  {\bibfnamefont{L.}~\bibnamefont{Benfatto}}, \bibinfo {author}
  {\bibfnamefont{M.}~\bibnamefont{Neto}}, \bibinfo {author}
  {\bibfnamefont{E.}~\bibnamefont{Choi}},\ and\ \bibinfo {author}
  {\bibfnamefont{T.}~\bibnamefont{Sasagawa}},\ }%
  \bibfield{journal}{%
  \bibinfo {journal} {Arxiv preprint arXiv:1006.1891}}%
   (\bibinfo {year} {2010})%
  \bibAnnoteFile{NoStop}{Raicevic10}%
\bibitem{Berry84}%
  \BibitemOpen
  \bibfield{author}{%
  \bibinfo {author} {\bibfnamefont{M.~V.}\ \bibnamefont{Berry}},\ }%
  \bibfield{journal}{%
  \bibinfo {journal} {Proc. R. Soc. A}\ }%
  \textbf{\bibinfo {volume} {392}},\ \bibinfo {pages} {45} (\bibinfo {year}
  {1984})%
  \bibAnnoteFile{NoStop}{Berry84}%
\bibitem{Nagaosa10AHE}%
  \BibitemOpen
  \bibfield{author}{%
  \bibinfo {author} {\bibfnamefont{N.}~\bibnamefont{Nagaosa}}, \bibinfo
  {author} {\bibfnamefont{J.}~\bibnamefont{Sinova}}, \bibinfo {author}
  {\bibfnamefont{S.}~\bibnamefont{Onoda}}, \bibinfo {author}
  {\bibfnamefont{A.~H.}\ \bibnamefont{MacDonald}},\ and\ \bibinfo {author}
  {\bibfnamefont{N.~P.}\ \bibnamefont{Ong}},\ }%
  \bibfield{journal}{%
  \bibinfo {journal} {Rev. Mod. Phys.}\ }%
  \textbf{\bibinfo {volume} {82}},\ \bibinfo {pages} {1539} (\bibinfo {year}
  {2010})%
  \bibAnnoteFile{NoStop}{Nagaosa10AHE}%
\bibitem{Ye99}%
  \BibitemOpen
  \bibfield{author}{%
  \bibinfo {author} {\bibfnamefont{J.}~\bibnamefont{Ye}}, \bibinfo {author}
  {\bibfnamefont{Y.~B.}\ \bibnamefont{Kim}}, \bibinfo {author}
  {\bibfnamefont{A.~J.}\ \bibnamefont{Millis}}, \bibinfo {author}
  {\bibfnamefont{B.~I.}\ \bibnamefont{Shraiman}}, \bibinfo {author}
  {\bibfnamefont{P.}~\bibnamefont{Majumdar}},\ and\ \bibinfo {author}
  {\bibfnamefont{Z.}~\bibnamefont{Te{\v {s}}anovi{\'{c}}}},\ }%
  \bibfield{journal}{%
  \bibinfo {journal} {Phys. Rev. Lett.}\ }%
  \textbf{\bibinfo {volume} {83}},\ \bibinfo {pages} {3737} (\bibinfo {year}
  {1999})%
  \bibAnnoteFile{NoStop}{Ye99}%
\bibitem{Matl98}%
  \BibitemOpen
  \bibfield{author}{%
  \bibinfo {author} {\bibfnamefont{P.}~\bibnamefont{Matl}}, \bibinfo {author}
  {\bibfnamefont{N.~P.}\ \bibnamefont{Ong}}, \bibinfo {author}
  {\bibfnamefont{Y.~F.}\ \bibnamefont{Yan}}, \bibinfo {author}
  {\bibfnamefont{Y.~Q.}\ \bibnamefont{Li}}, \bibinfo {author}
  {\bibfnamefont{D.}~\bibnamefont{Studebaker}}, \bibinfo {author}
  {\bibfnamefont{T.}~\bibnamefont{Baum}},\ and\ \bibinfo {author}
  {\bibfnamefont{G.}~\bibnamefont{Doubinina}},\ }%
  \bibfield{journal}{%
  \bibinfo {journal} {Phys. Rev. B}\ }%
  \textbf{\bibinfo {volume} {57}},\ \bibinfo {pages} {10248} (\bibinfo {year}
  {1998})%
  \bibAnnoteFile{NoStop}{Matl98}%
\bibitem{Taguchi01}%
  \BibitemOpen
  \bibfield{author}{%
  \bibinfo {author} {\bibfnamefont{T.}~\bibnamefont{Taguchi}}, \bibinfo
  {author} {\bibfnamefont{Y.}~\bibnamefont{Oohara}}, \bibinfo {author}
  {\bibfnamefont{H.}~\bibnamefont{Yoshizawa}}, \bibinfo {author}
  {\bibfnamefont{N.}~\bibnamefont{Nagaosa}},\ and\ \bibinfo {author}
  {\bibfnamefont{Y.}~\bibnamefont{Tokura}},\ }%
  \bibfield{journal}{%
  \bibinfo {journal} {Science}\ }%
  \textbf{\bibinfo {volume} {291}},\ \bibinfo {pages} {2573} (\bibinfo {year}
  {2001})%
  \bibAnnoteFile{NoStop}{Taguchi01}%
\bibitem{Lee09}%
  \BibitemOpen
  \bibfield{author}{%
  \bibinfo {author} {\bibfnamefont{M.}~\bibnamefont{Lee}}, \bibinfo {author}
  {\bibfnamefont{W.}~\bibnamefont{Kang}}, \bibinfo {author}
  {\bibfnamefont{Y.}~\bibnamefont{Onose}}, \bibinfo {author}
  {\bibfnamefont{Y.}~\bibnamefont{Tokura}},\ and\ \bibinfo {author}
  {\bibfnamefont{N.~P.}\ \bibnamefont{Ong}},\ }%
  \bibfield{journal}{%
  \bibinfo {journal} {Phys. Rev. Lett.}\ }%
  \textbf{\bibinfo {volume} {102}},\ \bibinfo {pages} {186601} (\bibinfo {year}
  {2009})%
  \bibAnnoteFile{NoStop}{Lee09}%
\bibitem{Neubauer09}%
  \BibitemOpen
  \bibfield{author}{%
  \bibinfo {author} {\bibfnamefont{A.}~\bibnamefont{Neubauer}}, \bibinfo
  {author} {\bibfnamefont{C.}~\bibnamefont{Pfleiderer}}, \bibinfo {author}
  {\bibfnamefont{B.}~\bibnamefont{Binz}}, \bibinfo {author}
  {\bibfnamefont{A.}~\bibnamefont{Rosch}}, \bibinfo {author}
  {\bibfnamefont{R.}~\bibnamefont{Ritz}}, \bibinfo {author}
  {\bibfnamefont{P.~G.}\ \bibnamefont{Niklowitz}},\ and\ \bibinfo {author}
  {\bibfnamefont{P.}~\bibnamefont{B\"oni}},\ }%
  \bibfield{journal}{%
  \bibinfo {journal} {Phys. Rev. Lett.}\ }%
  \textbf{\bibinfo {volume} {102}},\ \bibinfo {pages} {186602} (\bibinfo {year}
  {2009})%
  \bibAnnoteFile{NoStop}{Neubauer09}%
\bibitem{Timm09}%
  \BibitemOpen
  \bibfield{author}{%
  \bibinfo {author} {\bibfnamefont{C.}~\bibnamefont{Timm}},\ }%
  \bibfield{journal}{%
  \bibinfo {journal} {Physics}\ }%
  \textbf{\bibinfo {volume} {2}},\ \bibinfo {pages} {35} (\bibinfo {year}
  {2009})%
  \bibAnnoteFile{NoStop}{Timm09}%
\bibitem{Barnes07}%
  \BibitemOpen
  \bibfield{author}{%
  \bibinfo {author} {\bibfnamefont{S.~E.}\ \bibnamefont{Barnes}}\ and\ \bibinfo
  {author} {\bibfnamefont{S.}~\bibnamefont{Maekawa}},\ }%
  \bibfield{journal}{%
  \bibinfo {journal} {Phys. Rev. Lett.}\ }%
  \textbf{\bibinfo {volume} {98}},\ \bibinfo {pages} {246601} (\bibinfo {year}
  {2007})%
  \bibAnnoteFile{NoStop}{Barnes07}%
\bibitem{Hai09}%
  \BibitemOpen
  \bibfield{author}{%
  \bibinfo {author} {\bibfnamefont{P.}~\bibnamefont{Hai}}, \bibinfo {author}
  {\bibfnamefont{S.}~\bibnamefont{Ohya}}, \bibinfo {author}
  {\bibfnamefont{M.}~\bibnamefont{Tanaka}}, \bibinfo {author}
  {\bibfnamefont{S.}~\bibnamefont{Barnes}},\ and\ \bibinfo {author}
  {\bibfnamefont{S.}~\bibnamefont{Maekawa}},\ }%
  \bibfield{journal}{%
  \bibinfo {journal} {Nature}\ }%
  \textbf{\bibinfo {volume} {458}},\ \bibinfo {pages} {489} (\bibinfo {year}
  {2009})%
  \bibAnnoteFile{NoStop}{Hai09}%
\bibitem{Yang09}%
  \BibitemOpen
  \bibfield{author}{%
  \bibinfo {author} {\bibfnamefont{S.~A.}\ \bibnamefont{Yang}}, \bibinfo
  {author} {\bibfnamefont{G.~S.~D.}\ \bibnamefont{Beach}}, \bibinfo {author}
  {\bibfnamefont{C.}~\bibnamefont{Knutson}}, \bibinfo {author}
  {\bibfnamefont{D.}~\bibnamefont{Xiao}}, \bibinfo {author}
  {\bibfnamefont{Q.}~\bibnamefont{Niu}}, \bibinfo {author}
  {\bibfnamefont{M.}~\bibnamefont{Tsoi}},\ and\ \bibinfo {author}
  {\bibfnamefont{J.~L.}\ \bibnamefont{Erskine}},\ }%
  \bibfield{journal}{%
  \bibinfo {journal} {Phys. Rev. Lett.}\ }%
  \textbf{\bibinfo {volume} {102}},\ \bibinfo {pages} {067201} (\bibinfo {year}
  {2009})%
  \bibAnnoteFile{NoStop}{Yang09}%
\bibitem{Note1}%
  \BibitemOpen
  \bibinfo {note} {See Ref.~\cite {Rossler06} for discussions about $\protect
  \mathcal {H}_{\protect \bf m}$.}%
  \bibAnnoteFile{Stop}{Note1}%
\bibitem{Korenman77}%
  \BibitemOpen
  \bibfield{author}{%
  \bibinfo {author} {\bibfnamefont{V.}~\bibnamefont{Korenman}}, \bibinfo
  {author} {\bibfnamefont{J.~L.}\ \bibnamefont{Murray}},\ and\ \bibinfo
  {author} {\bibfnamefont{R.~E.}\ \bibnamefont{Prange}},\ }%
  \bibfield{journal}{%
  \bibinfo {journal} {Phys. Rev. B}\ }%
  \textbf{\bibinfo {volume} {16}},\ \bibinfo {pages} {4032} (\bibinfo {year}
  {1977})%
  \bibAnnoteFile{NoStop}{Korenman77}%
\bibitem{Volovik87}%
  \BibitemOpen
  \bibfield{author}{%
  \bibinfo {author} {\bibfnamefont{G.}~\bibnamefont{Volovik}},\ }%
  \bibfield{journal}{%
  \bibinfo {journal} {J. Phys. C}\ }%
  \textbf{\bibinfo {volume} {20}},\ \bibinfo {pages} {L83} (\bibinfo {year}
  {1987})%
  \bibAnnoteFile{NoStop}{Volovik87}%
\bibitem{Tserkovnyak08}%
  \BibitemOpen
  \bibfield{author}{%
  \bibinfo {author} {\bibfnamefont{Y.}~\bibnamefont{Tserkovnyak}}\ and\
  \bibinfo {author} {\bibfnamefont{M.}~\bibnamefont{Mecklenburg}},\ }%
  \bibfield{journal}{%
  \bibinfo {journal} {Phys. Rev. B}\ }%
  \textbf{\bibinfo {volume} {77}},\ \bibinfo {pages} {134407} (\bibinfo {year}
  {2008})%
  \bibAnnoteFile{NoStop}{Tserkovnyak08}%
\bibitem{Wong09}%
  \BibitemOpen
  \bibfield{author}{%
  \bibinfo {author} {\bibfnamefont{C.~H.}\ \bibnamefont{Wong}}\ and\ \bibinfo
  {author} {\bibfnamefont{Y.}~\bibnamefont{Tserkovnyak}},\ }%
  \bibfield{journal}{%
  \bibinfo {journal} {Phys. Rev. B}\ }%
  \textbf{\bibinfo {volume} {80}},\ \bibinfo {pages} {184411} (\bibinfo {year}
  {2009})%
  \bibAnnoteFile{NoStop}{Wong09}%
\bibitem{Clem91}%
  \BibitemOpen
  \bibfield{author}{%
  \bibinfo {author} {\bibfnamefont{J.~R.}\ \bibnamefont{Clem}},\ }%
  \bibfield{journal}{%
  \bibinfo {journal} {Phys. Rev. B}\ }%
  \textbf{\bibinfo {volume} {43}},\ \bibinfo {pages} {7837} (\bibinfo {year}
  {1991})%
  \bibAnnoteFile{NoStop}{Clem91}%
\bibitem{Ohe09}%
  \BibitemOpen
  \bibfield{author}{%
  \bibinfo {author} {\bibfnamefont{J.-i.}\ \bibnamefont{Ohe}}, \bibinfo
  {author} {\bibfnamefont{S.~E.}\ \bibnamefont{Barnes}}, \bibinfo {author}
  {\bibfnamefont{H.-W.}\ \bibnamefont{Lee}},\ and\ \bibinfo {author}
  {\bibfnamefont{S.}~\bibnamefont{Maekawa}},\ }%
  \bibfield{journal}{%
  \bibinfo {journal} {Appl. Phys. Lett.}\ }%
  \textbf{\bibinfo {volume} {95}},\ \bibinfo {pages} {123110} (\bibinfo {year}
  {2009})%
  \bibAnnoteFile{NoStop}{Ohe09}%
\bibitem{Yu10}%
  \BibitemOpen
  \bibfield{author}{%
  \bibinfo {author} {\bibfnamefont{X.}~\bibnamefont{Yu}}, \bibinfo {author}
  {\bibfnamefont{Y.}~\bibnamefont{Onose}}, \bibinfo {author}
  {\bibfnamefont{N.}~\bibnamefont{Kanazawa}}, \bibinfo {author}
  {\bibfnamefont{J.}~\bibnamefont{Park}}, \bibinfo {author}
  {\bibfnamefont{J.}~\bibnamefont{Han}}, \bibinfo {author}
  {\bibfnamefont{Y.}~\bibnamefont{Matsui}}, \bibinfo {author}
  {\bibfnamefont{N.}~\bibnamefont{Nagaosa}},\ and\ \bibinfo {author}
  {\bibfnamefont{Y.}~\bibnamefont{Tokura}},\ }%
  \bibfield{journal}{%
  \bibinfo {journal} {Nature}\ }%
  \textbf{\bibinfo {volume} {465}},\ \bibinfo {pages} {901} (\bibinfo {year}
  {2010})%
  \bibAnnoteFile{NoStop}{Yu10}%
\bibitem{Wang06}%
  \BibitemOpen
  \bibfield{author}{%
  \bibinfo {author} {\bibfnamefont{Y.}~\bibnamefont{Wang}}, \bibinfo {author}
  {\bibfnamefont{L.}~\bibnamefont{Li}},\ and\ \bibinfo {author}
  {\bibfnamefont{N.~P.}\ \bibnamefont{Ong}},\ }%
  \bibfield{journal}{%
  \bibinfo {journal} {Phys. Rev. B}\ }%
  \textbf{\bibinfo {volume} {73}},\ \bibinfo {pages} {024510} (\bibinfo {year}
  {2006})%
  \bibAnnoteFile{NoStop}{Wang06}%
\bibitem{Note2}%
  \BibitemOpen
  \bibinfo {note} {Recent magnetization and magnetoresistance measurements\cite
  {Raicevic10} suggested the presence of skyrmions in insulating
  antiferromagnet La$_2$Cu$_{0.97}$Li$_{0.03}$O$_4$.}%
  \bibAnnoteFile{Stop}{Note2}%
\end{thebibliography}%

\end{document}